\documentclass[reqno]{amsart}
 \usepackage{graphicx}          %
 \usepackage{amssymb}
 \usepackage{amsmath}
\usepackage{cleveref}
\usepackage{algpseudocode}

 \usepackage{calc}
 \usepackage{tikz}
 \usepackage{pgfplots}
\usepackage{cases}

\usetikzlibrary{arrows}
\usetikzlibrary{shapes}
 \usetikzlibrary{calc}
\usetikzlibrary{decorations.pathreplacing}
\usetikzlibrary{arrows,positioning} 
\usetikzlibrary{pgfplots.groupplots}

\usepackage{thmtools}
\newtheorem{rem}{Remark}

\usepackage{tikz}
\usetikzlibrary{arrows.meta} 
\usepackage{pgfplots}
\usetikzlibrary{calc}
\usepackage{ushort}
\usetikzlibrary{arrows}
\usetikzlibrary{shapes}

\tikzstyle{link}=[line width=2pt, ->,>=latex]
\tikzstyle{junc}=[draw,circle,inner sep=1pt,minimum width=8pt]
\tikzstyle{onramp}=[line width=2pt, dotted,->,>=latex]
\usepackage{mathtools}

\usepackage{stmaryrd}

\begin{document}

\title{A Benchmark Problem in Transportation Networks}

\author{Samuel Coogan and Murat Arcak}
\thanks{S. Coogan is with the School of Electrical and Computer Engineering and the School of Civil and Environmental Engineering at the Georgia Institute of Technology, Atlanta, GA. \texttt{sam.coogan@gatech.edu}. M. Arcak is with the Department of Electrical Engineering and Computer Sciences, University of California, Berkeley, Berkeley, CA. \texttt{arcak@eecs.berkeley.edu}}
\maketitle
\makeatletter
\newcommand{\pushright}[1]{\ifmeasuring@#1\else\omit\hfill$\displaystyle#1$\fi\ignorespaces}
\newcommand{\pushleft}[1]{\ifmeasuring@#1\else\omit$\displaystyle#1$\hfill\fi\ignorespaces}
\makeatother

\begin{abstract}
In this note, we propose a case study of freeway traffic flow modeled as a hybrid system. We describe two general classes of networks that model flow along a freeway with merging onramps. The admission rate of traffic flow from each onramp is metered via a control input. Both classes of networks are easily scaled to accommodate arbitrary state dimension. The model is discrete-time and possesses piecewise-affine dynamics. Moreover, we present several control objectives that are especially relevant for traffic flow management. The proposed model is flexible and extensible and offers a benchmark for evaluating tools and techniques developed for hybrid systems.
\end{abstract}

\section{Introduction}
\label{sec:introduction}

Traffic flow theory has its foundations in the \emph{Lighthill-Whitham-Richards (LWR) model}, a first-order partial differential equation in the form of a conservation law that models traffic flow on a single road \cite{Lighthill:1955vn, Richards:1956ys}.  Originally motivated by the need to efficiently simulate traffic flow, the \emph{cell transmission model} was proposed as a finite-dimensional approximation to the LWR model \cite{Daganzo:1995kx, Lebacque:1996zr}. However, the cell transmission model has since been established as an appropriate model of traffic flow in its own right \cite{Gomes:2006uq, Gomes:2008fk, kurzhanskiy2010active}, and recent research has focused on studying the dynamical properties of this model \cite{coogan2015compartmental, Lovisari:2014qv, Wright:2016nx}.

The cell transmission model considers traffic networks as interconnected links or compartments with finite capacity to store vehicles. Vehicles flow from link to link over time, and thus the \emph{occupancy} of a link is time-varying. The cell transmission model adopts a fluid approximation of traffic flow so that occupancy is not restricted to integer values.  Flow of vehicles from an upstream link to a downstream link is restricted by the \emph{demand} of vehicles on the upstream link to flow downstream, and the \emph{supply} of capacity available on the downstream link to accept incoming flow. These restrictions give rise to a hybrid model as the behavior is different in the supply-restricted regime. For junctions with multiple incoming links or multiple outgoing links, demand is divided among the outgoing links and supply is divided among the incoming links, and a variety of specific models for this division has been proposed in the literature; see the above-cited references.

In this note, we present a particular instantiation of the cell transmission model especially amenable to analysis and control as a hybrid system. We limit our attention to models where each junction has a single incoming link and at most two outgoing links (diverging junction), or at most two incoming links and a single outgoing link (merging junction), and the result is a piecewise-affine traffic flow model.

The purpose of this note is to distill existing traffic models into a simple but extensible model of traffic flow and offer it as a practically motivated case study in need of computationally efficient and scalable tools for control verification and synthesis. The model presented here agrees with the various models presented in the literature and cited above and illuminates the fundamental challenges. 

The note is organized as follows. In Section \ref{sec:traffic-flow-model}, we present models for traffic flow at merging and diverging junctions. These simple models can be interconnected to create traffic networks with arbitrary topology. In Section \ref{sec:two-netw-topol}, we suggest two particular classes of network topologies and provide detailed models for both. In Section \ref{sec:perf-spec}, we present several control objectives for these two classes of networks which are especially relevant for traffic networks. In Section \ref{sec:discussion}, we identify several properties of the traffic flow model and make connections to existing results in the literature. Section \ref{sec:conclusions} contains concluding remarks.

\section{Traffic Flow at Junctions}
\label{sec:traffic-flow-model}

We begin by discussing the two elemental traffic flow models: the \emph{merging} junction where two traffic links merge to one downstream link as in Figure \ref{fig:mandd}(a), and the \emph{diverging} junction where one link diverges to two downstream links as in Figure \ref{fig:mandd}(b). These two junction models capture the essential dynamics exhibited by traffic flow networks. In Section \ref{sec:two-netw-topol}, we combine these models and propose two benchmark network topologies. While this note considers networks with only merging nodes and diverging nodes, the model is easily extended to accommodate nodes with multiple incoming and outgoing links \cite{Coogan:2014pi}.

Let $x_i[t]$ denote the number of vehicles on link $i$ at time $t$, that is, $x_i$ is the \emph{occupancy} of link $i$. We adopt a macroscopic modeling approach and assume that $x_i[t]$ takes continuous values. The state of a traffic flow network is the collection of link occupancies in the network so that, for the merging and the diverging junctions, $x[t]=\begin{pmatrix} x_1[t] & x_2[t] & x_3[t]\end{pmatrix}^T$ is the system state.

The flow of traffic from link to link through merging and diverging junctions is a function of the number of vehicles on a link wishing to flow downstream, which is called the \emph{demand} of the link, the available downstream road space to accommodate these vehicles, which is called the \emph{supply} of the link, and possibly a control input, which is discussed below.  Link demand is an increasing function in the number of vehicles on the link, and the link supply is a decreasing function in the number of vehicles on the link. In the transportation literature, the demand and supply functions are referred to as the \emph{fundamental diagram} relating link occupancy to flow.

A common approach is to adopt a triangular fundamental diagram so that the demand and supply functions are affine in $x_i$ with the additional restriction that the demand saturates at some maximum value. Thus we model the demand function $D(x)$ and the supply function $S(x)$ as
\begin{align}
\label{eq:dem}
  D(x)&=\min\{c,v x\}\\
    S(x)&=w(\bar{x}-x)
\end{align}
where $c$ is the \emph{capacity}, $v$ is the \emph{free-flow speed}, $w$ is the \emph{congestion-wave speed}, and $\bar{x}$ is the \emph{jam occupancy} of a link \cite{Gomes:2008fk}. Standard values for these parameters are given in Table \ref{tab:params}, which are appropriate for links modeling two lanes of traffic, with each link one mile long, and a time step of 30 seconds \cite{Gomes:2008fk}.

\begin{table}
  \centering
  \begin{tabular}{r l l}
    Paramter &Value&Units\\
\hline \hline
    Link length&1& mile\\
    Period&0.5& min\\
    $c$&40& veh$/$period\\
    $v$&0.5& links$/$period \\
    $w$&${0.5}/{3}$&links$/$period\\
    $\bar{x}$&320&vehicles\\
    $\beta$&0.75&\multicolumn{1}{c}{---}\\
    $\alpha$&1&\multicolumn{1}{c}{---}\\
    $\bar{\alpha}$&5&\multicolumn{1}{c}{---}
  \end{tabular}
  \caption{Parameter values}
  \label{tab:params}
\end{table}

  We remark that it is common to consider different parameters for different links to accommodate, \emph{e.g.}, varying road geometry, varying link lengths, \emph{etc.} For notational convenience and to establish a consistent model, we assume all parameters take the values shown in Table \ref{tab:params} for all links.

In traffic networks, link flow may be artificially restricted, or \emph{metered}, via traffic signaling devices, and, in this note, metering is considered to be the only available control input.  For example, metering is commonly encountered on freeway onramps to control admittance to the freeway. In particular, if a link is metered via control input $u$, then the traffic flow that exits the link cannot exceed $u[t]$ at time $t$.

The fundamental rule governing the dynamics of traffic networks is that the flow exiting a link should not exceed the link's demand, nor the link's metering rate if the link is controlled, and it should also not exceed downstream supply. Thus, for a junction consisting of one incoming link, labeled link $1$, and one outgoing link, labeled link $2$, the state equations are
\begin{align}
  \label{eq:4}
  x_1[t+1]&=x_1[t]-\min\{D(x_1[t]),S(x_2[t])\}+d[t],\\
  x_2[t+1]&=x_2[t]+\min\{D(x_1[t]),S(x_2[t])\} - D(x_2[t]),
\end{align}
where $x_i[t]$ is the state, \emph{i.e.}, occupancy, of link $i\in\{1,2\}$ at time $t$, $d[t]$ is the number of vehicles arriving on link $1$ from upstream. We assume there are no links downstream of link $2$ so that the flow exiting link $2$ is equal to demand.

 We now extend this rule in a natural way for the merging junction and the diverging junction shown in Figure \ref{fig:mandd}.

\subsection{Merge Junction}

\begin{figure}
  \centering
\begin{tabular}{c c}
  \begin{tikzpicture}[scale=.7]
\node[junc] (j1) at (0,0) {};    
\draw[link] ($(j1)+(180:.8in)$)--node[above]{$1$} (j1);
\draw[onramp] ($(j1)+(210:.8in)$)--node[below]{$2$} (j1);
\draw[link] (j1)--node[below]{$3$} ($(j1)+(0:.8in)$);
  \end{tikzpicture}&
  \begin{tikzpicture}[scale=.7]
\node[junc] (j1) at (0,0) {};    
\draw[link] ($(j1)+(180:.8in)$)--node[above]{$1$} (j1);
\draw[link] (j1)--node[above]{$2$} ($(j1)+(30:.8in)$);
\draw[link] (j1)--node[below]{$3$} ($(j1)+(-30:.8in)$);
  \end{tikzpicture}\\
(a)&(b)
\end{tabular}
\vspace{-.1in}
\caption{(a) At a \emph{merging junction}, traffic flow from two incoming links merge and flow downstream to an outgoing link. It is common for one of the incoming links (dashed) to model an \emph{onramp} for which the rate of traffic flow can be controlled using signaling devices, thus providing a control input to the system. (b) At a \emph{diverging junction}, traffic flows from one incoming link to two outgoing links. Diverging junctions exhibit the \emph{first-in-first-out} property whereby congestion (lack of supply) on one outgoing link reduces flow to the other outgoing link.}
  \label{fig:mandd}
\end{figure}
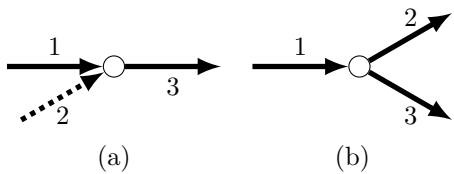

At the merging junction show in Figure \ref{fig:mandd}(a), links $1$ and $2$ flow downstream to link $3$. In this note, merging junctions are interpreted as modeling a freeway entrance ramp that joins a freeway, and thus links $1$ and $3$ are the freeway and link $2$ is the onramp. Link $2$ is metered via control input $u$. Let $d_1[t]$ and $d_2[t]$ be the exogenous, uncontrolled arrival of vehicles to links $1$ and $2$, respectively. A fraction $\beta\leq 1$ of vehicles exiting link $1$ join link $3$, and the remaining $1-\beta$ fraction are assumed to exit the network via an unmodeled exit link. All vehicles flowing from the onramp link $2$ join link $3$. 

Moreover, it is assumed that the supply of link $3$ available to link $1$  (respectively, link $2$) is determined by a fixed weight $\alpha>0$ (respectively, $\bar{\alpha}>0$). The values for $\beta$, $\alpha$, and $\bar{\alpha}$ used in this note are reported in Table \ref{tab:params}. 

The state-update equations for the merge junction are given by

\begin{align}
\label{eq:m1}
  x_1[t+1]&= x_1[t]-\min\left\{ D(x_1[t]),\frac{\alpha}{\beta} S(x_3[t])\right\}+d_1[t],\\
  x_{2}[t+1]&=x_{2}[t]-\min\left\{ D(x_{2}[t]),\bar{\alpha}S(x_3[t]), u[t]\right\}+d_2[t],\\
\nonumber   x_3[t+1]&=x_3[t]-D(x_3[t])+\min\left\{ \beta D(x_1[t]),{\alpha}S(x_3[t])\right\}\\
\label{eq:m2}&\quad +\min\left\{D(x_{2}[t]),{\bar{\alpha}}S(x_3[t]), u[t]\right\}.
\end{align}

\begin{rem}
Note that $\frac{1}{w}S(x_i[t])=(\bar{x}-x_i[t])$ is the available space on link $i$ at time $t$ so that we interpret $w\alpha$ (respectively, $w\bar{\alpha}$) as the fraction of this space available to link $1$ (respectively, link $2$). Thus, occupancy will not exceed $\bar{x}$ so long as $w\alpha+w\bar{\alpha}\leq 1$, which holds for the values in Table \ref{tab:params}. The model is referred to as an \emph{asymmetrical} cell transmission model since $\bar{\alpha}>1$ and therefore flow from link $2$ may exceed supply \cite{Gomes:2006uq, Gomes:2008fk}.

\end{rem}

\subsection{Diverge Junction}
A diverge junction models the division of flow from one incoming link to two outgoing links. We assume that the outgoing flow of link $1$ in Figure \ref{fig:mandd}(b) divides evenly among the outgoing links $2$ and $3$.
 We do not assume any of the links in the diverging junction are metered.

Diverging junctions in traffic flow networks have been empirically observed to obey a \emph{first-in-first-out (FIFO)} property whereby congestion on one outgoing link restricts flow to the other outgoing link, that is, lack of supply on link $2$ restricts flow to link $3$ and vice-versa. The intuition for this phenomenon is that traffic waiting to move downstream to link $2$ blocks traffic destined for link $3$, even though link $3$ has adequate supply. This intuition has been empirically observed to extend to traffic flow on multi-lane freeways \cite{Munoz:2002qv}. In the present model, we assume a \emph{full FIFO} property whereby complete congestion on one outgoing link completely restricts flow to other outgoing links. This is a common assumption in transportation flow models and can be relaxed to obtain partial FIFO models \cite{Coogan:2016rp}.

The state-update equations for the diverge junction are given by
\begin{align}
\label{eq:d1} x_1[t+1]&=x_1[t]-\min\left\{D(x_1[t]),2S(x_{2}[t]),2S(x_{3}[t])\right\}+d_1[t],\\
x_{2}[t+1]&=x_{2}[t]+\min\left\{0.5D(x_1[t]),S(x_{2}[t]),S(x_{3}[t])\right\}-D(x_{2}[t]),\\
\label{eq:d3} x_{3}[t+1]&=x_{3}[t]+\min\left\{0.5D(x_1[t]),S(x_{2}[t]),S_{3}(x[t])\right\}-D(x_{3}[t]).
\end{align}

\begin{figure*}
  \centering
  \begin{alignat}{2}
x_1[t+1]&=x_1[t]-\min\left\{D(x_1[t]),\frac{\alpha}{\beta} S(x_{2}[t])\right\}+d_1[t],\\
\nonumber x_N[t+1]=&x_N[t]-D(x_N[t])+\min\left\{ \beta D(x_{N-1}[t]),{\alpha}S(x_N[t])\right\}\\
&+\min\left\{D(x_{(N-1)'}[t]),{\bar{\alpha}}S(x_N[t]), u_{(N-1)'}[t]\right\},\\
\nonumber x_i[t+1]=&x_i[t]-\min\left\{D(x_i[t]),\frac{\alpha}{\beta} S(x_{i+1}[t])\right\}+\min\left\{\beta D(x_{i-1}[t]),\alpha S(x_i[t])\right\}\\
&+\min\left\{D(x_{(i-1)'}[t]),\bar{\alpha}S(x_i[t]),u_{(i-1)'}[t]\right\} \qquad {\text{for }  i=2,\ldots,N-1,\qquad}&\\
\nonumber    x_{i'}[t+1]=&x_{i'}[t]-\min\left\{D(x_{i'}[t]),{\bar{\alpha}}S(x_{i+1}[t]),u_{i'}[t]\right\}+d_{i'}[t]\\
&{\text{for }i'=1',\ldots,(N-1)'}.\qquad&
  \end{alignat}
  \caption{State-update equations for the simple freeway.}
  \label{fig:netw1}
\end{figure*}

\section{Two Network Topologies}
\label{sec:two-netw-topol}
We now use the merge and diverge junction models presented in Section \ref{sec:traffic-flow-model} as building blocks to construct two classes of traffic networks. These two classes provide an extensible benchmark problem for analysis and control of hybrid systems.
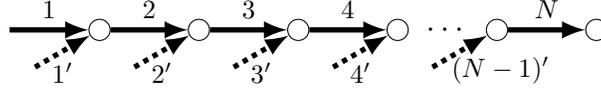
\begin{figure}
  \centering
        \begin{tikzpicture}[scale=.65]
\node (j0) at (0,0) {};
\node[junc] (j1) at ($(j0)+(0:.8in)$) {};
\node[junc] (j2) at ($(j1)+(0:.8in)$) {};
\node[junc] (j3) at ($(j2)+(0:.8in)$) {};
\node[junc] (j4) at ($(j3)+(0:.8in)$) {};
\node at ($(j4)+(0:.4in)$) {\Large $\cdots$};
\node[junc] (j5) at ($(j4)+(0:.8in)$) {};
\node[junc] (j6) at ($(j5)+(0:.8in)$) {};
\draw[link] (j0)-- node[above]{$1$}(j1);
\draw[link] (j1)--node[above]{$2$} (j2);
\draw[link] (j2)--node[above]{$3$} (j3);
\draw[link] (j3)--node[above]{$4$} (j4);
\draw[link] (j5)--node[above]{$N$} (j6);

\draw[onramp] ($(j1)+(210:.6in)$)--node[below]{$1'$} (j1);
\draw[onramp] ($(j2)+(210:.6in)$)--node[below]{$2'$} (j2);
\draw[onramp] ($(j3)+(210:.6in)$)--node[below]{$3'$} (j3);
\draw[onramp] ($(j4)+(210:.6in)$)--node[below]{$4'$} (j4);
\draw[onramp] ($(j5)+(210:.6in)$)--node[right=5pt,pos=-.1]{$(N-1)'$} (j5);
    \end{tikzpicture}
\vspace{-.1in}
  \caption{A \emph{simple freeway network} consists of only merging junctions and models a length of freeway with onramps. The length-$N$ simple freeway consists of $(2N-1)$ links, $(N-1)$ of which model \emph{onramps} and thus possess control inputs. Therefore, the state dimension of the network is $(2N-1)$, and the input dimension is $(N-1)$. }
  \label{fig:network1}
  \end{figure}

\subsection{Simple Freeway Network}
\label{sec:two-networks}
The first class of benchmark networks consists of only merging junctions and models a length of freeway with onramps. This class of network topologies is commonly encountered in studies of ramp metering for freeways \cite{Gomes:2006uq,Gomes:2008fk, Reilly:2015zr}. Given parameter $N$, the \emph{length-$N$ simple freeway} is composed of $2N-1$ freeway links, $N-1$ of which model \emph{onramps} and thus possess control inputs, as shown in Figure \ref{fig:network1}. The state dimension of the network is $(2N-1)$ and the input dimension is $(N-1)$.

The state-update equations for the length-$N$ simple freeway are given in Figure \ref{fig:netw1}.

\begin{rem}
  The simple freeway model may be modified to arrive at various alternative configurations. For example, for fixed $N$, the number of states and inputs may be reduced by eliminating some onramps from the model.
\end{rem}

\subsection{Diverging Freeway Network}
\label{sec:diverg-freew-netw}
\begin{figure}
  \centering
{\small
    \begin{tikzpicture}[scale=.6]
      \node[junc] (j0) at (0,0) {};
     \node[junc] (jn1) at ($(j0)+(-.8in,0)$) {};
     \node[junc] (jn2) at ($(jn1)+(-.8in,0)$) {};
      \draw[link] ($(jn2)+(180:.8in)$) to node[above]{$-M$}(jn2);
      \draw[link] (jn1) to node[above]{$0$} (j0);
      \draw[onramp] ($(jn1)+(-135:.8in)$) --node[right=2pt, pos=.1]{$-1'$} (jn1);
      \draw[onramp] ($(jn2)+(-135:.8in)$) --node[right=2pt, pos=.1]{$-M'$} (jn2);
            \node at ($(jn2)+(0:.4in)$) {\Large $\cdots$};

      \node[junc] (j1a) at ($(j0)+(45:.8in)$) {};
      \node[junc] (j1b) at ($(j0)+(-45:.8in)$) {};
      \draw[link] (j0) -- node[left=3pt,pos=.8]{$1$} (j1a);
      \draw[link] (j0) -- node[left=3pt,pos=.8]{$N+1$} (j1b);
      \draw[onramp] ($(j1a)+(135:.8in)$) --node[right=2pt, pos=.1]{$1'$} (j1a);
      \draw[onramp] ($(j1b)+(-135:.8in)$) --node[right, pos=.1]{\tiny $(N\hspace{-2pt}+\hspace{-2pt}1)'$}  (j1b);
      \node[junc] (j1aa) at ($(j1a)+(0:.8in)$) {};
      \draw[onramp] ($(j1aa)+(135:.8in)$) --node[right=2pt, pos=.1]{$2'$} (j1aa);
      \node[junc] (j1bb) at ($(j1b)+(0:.8in)$) {};
      \draw[onramp] ($(j1bb)+(-135:.8in)$) --node[right=0pt, pos=.1]{\tiny $(N\hspace{-2pt}+\hspace{-2pt}2)'$} (j1bb);
      \draw[link] (j1a) -- node[below,pos=.5] {$2$} (j1aa);
      \draw[link] (j1b) -- node[above] {$N+2$} (j1bb);
      \node[junc] (j3) at ($(j1aa)+(0:.8in)$) {};
      \node[junc] (j5) at ($(j1bb)+(0:.8in)$) {};
      \node[junc] (j4) at ($(j3)+(0:.8in)$) {};
      \node[junc] (j6) at ($(j5)+(0:.8in)$) {};

      \draw[onramp] ($(j3)+(135:.8in)$) --node[right=2pt, pos=.1]{\tiny$(N\hspace{-2pt}-\hspace{-2pt}1)'$} (j3);
      \draw[onramp] ($(j5)+(-135:.8in)$) --node[right=-1pt, pos=.1]{\tiny $(2N\hspace{-2pt}-\hspace{-2pt}1)'$} (j5);
      \node at ($(j1aa)+(0:.4in)$) {\Large $\cdots$};
      \node at ($(j1bb)+(0:.4in)$) {\Large $\cdots$};
      \draw[link] (j3) to node[below]{$N$}(j4);
      \draw[link] (j5) to node[above]{$2N$}(j6);
    \end{tikzpicture}
\vspace{-.2in}
  }
  \caption{A \emph{diverging freeway network} models a freeway that diverges to two freeways. The \emph{length-$(M,N)$} diverging freeway is composed of  $2M+4N-1$ links, $M+2(N-1)$ of which model \emph{onramps} and thus possess control inputs.}
\label{fig:network2}
\end{figure}
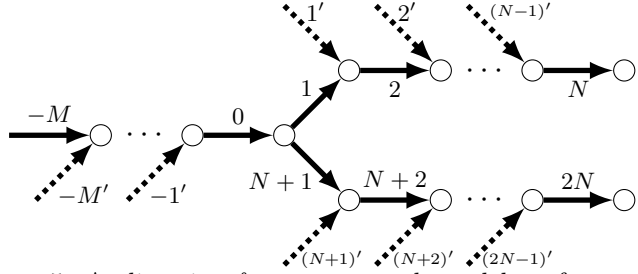

The second class of benchmark networks models a freeway that diverges into two freeways. Given parameters $M$ and $N$, the \emph{length-$(M,N)$ diverging freeway} is composed of  $2M+4N-1$ links, $M+2(N-1)$ of which model \emph{onramps} and thus possess control inputs, as shown in Figure \ref{fig:network2}. %

The state-update equations for the length-$N$ diverging freeway are given in Figure \ref{fig:netw2}.

\begin{figure*}
  \centering
 \begin{alignat}{2}
x_{(-M)}[t+1]&=x_{(-M)}[t]-\min\left\{D(x_{(-M)}[t]),\frac{\alpha}{\beta} S(x_{(-M+1)}[t])\right\}+d_{(-M)}[t],\\
\nonumber x_0[t+1]&=x_0[t]-\min\left\{D(x_0[t]),2S(x_{1}[t]),2S(x_{N+1}[t])\right\}\\
\nonumber&\quad +\min\left\{\beta D(x_{-1}[t]),\alpha S(x_0[t])\right\}\\
&\quad+\min\left\{D(x_{-1'}[t]),\bar{\alpha}S(x_0[t]),u_{-1'}[t]\right\},\\
\nonumber x_i[t+1]&=x_i[t]-\min\left\{ D(x_i[t]),\frac{\alpha}{\beta} S(x_{i+1}[t])\right\}\\
&\quad+\min\left\{0.5D(x_0[t]),S(x_{1}[t]),S(x_{N+1}[t])\right\}\text{for }i=1,N+1,\\
\nonumber x_i[t+1]&=x_i[t]-\min\left\{D(x_i[t]),\frac{\alpha}{\beta} S(x_{i+1}[t])\right\}+\min\left\{\beta D(x_{i-1}[t]),\alpha S(x_i[t])\right\}\\
\nonumber&\quad +\min\left\{D(x_{(i-1)'}[t]),\bar{\alpha}S(x_i[t]),u_{(i-1)'}[t]\right\}\\
&\quad\qquad \text{for }i=-M+1,\ldots,-1,2,\ldots,N-1,N+2,\ldots,2N-1, &\\
\nonumber x_i[t+1]&=x_i[t+1]+\min\left\{\beta D(x_{i-1}[t]),\alpha S(x_i[t])\right\}\\
\nonumber &\quad +\min\left\{D(x_{(i-1)'}[t]),\bar{\alpha}S(x_i[t]),u_{(i-1)'}[t]\right\}-D(x_i[t])\\
&\quad \qquad \text{for }i=N, 2N\\
\nonumber x_{i'}[t+1]&=x_{i'}[t]-\min\left\{D(x_{i'}[t]),{\bar{\alpha}}S(x_{i+1}[t]),u_{i'}[t]\right\}+d_{i'}[t]\\
\nonumber &\quad \qquad\text{for }i'=-M',\ldots,-1',1',2',\ldots,(N-1)' ,\\
&\quad \qquad\qquad\quad  (N+1)',\ldots,(2N-1)'.  &
 \end{alignat}
  \caption{State-update equations for the length-$(M,N)$ diverging freeway. }
  \label{fig:netw2}
\end{figure*}

\begin{rem}
As with the previous benchmark network, the diverging freeway model may be easily modified by eliminating onramps, which reduces the number of states and inputs of the model.
\end{rem}

\begin{rem}
The incoming flows to the onramps and to the first link of the network (\emph{i.e.}, link $1$ for the simple freeway and link $-M$ for the diverging freeway) are not subject to supply restrictions. The motivation for doing so is that the model is able to accommodate arbitrary exogenous flows as discussed in Section \ref{sec:exogenous-inputs}.
\end{rem}

\subsection{Exogenous Inputs}
\label{sec:exogenous-inputs}
For the simple freeway model, let
\begin{align}
  d[t]=\left(d_1[t],d_{1'}[t],d_{2'}[t],\ldots,d_{(N-1)'}[t]\right)\in\mathbb{R}^{N},
\end{align}
and, for the diverging freeway model, let
\begin{align}
\nonumber  d[t]=\left(\right.\hspace{-2pt}&d_{(-M)}[t],d_{(-M)'}[t],d_{(-M+1)'},\ldots , d_{(-1)'},\\
\nonumber &d_{1'}[t],\ldots,d_{(N-1)'}[t],d_{(N+1)'}[t],\ldots, d_{(2N-1)'}[t]\left.\hspace{-2pt}\right)\\
&\pushright{\in\mathbb{R}^{M+2N-1}\quad  }
\end{align}
  so that $d[t]$ is the vector of exogenous flows into the network at time $t$. 
Similarly, for the simple freeway, let 
\begin{align}
  u[t]=    \left(u_{1'}[t],u_{2'}[t],\ldots,u_{(N-1)'}[t]\right)\in \mathbb{R}^{N-1}
\end{align}
and, for the diverging freeway, let
\begin{align}
\nonumber u[t]=\left(\right.\hspace{-2pt}&u_{(-M')}[t],\ldots,u_{(-1)'}, u_{1'}[t],\ldots,u_{(N-1)'}[t],\\
&u_{(N+1)'}[t],\ldots,u_{(2N-1)'}[t]\left.\hspace{-2pt}\right)\qquad \in \mathbb{R}^{M+2(N-1)}
\end{align}  
so that $u[t]$ is the vector of inputs for the network at time $t$. Furthermore, let
\begin{align}
  \label{eq:3}
  n&=
  \begin{cases}
    2N-1&\text{for the simple freeway,}\\
2M+4N-1&\text{for the diverging freeway,}
  \end{cases}\\
m&=
  \begin{cases}
    N-1&\text{for the simple freeway,}\\
M+2(N-1)&\text{for the diverging freeway,}
  \end{cases} \quad \\
q&=
  \begin{cases}
    N&\text{for the simple freeway,}\\
M+2N-1&\text{for the diverging freeway}
  \end{cases}
\end{align}
be the state, input, and disturbance dimensions, respectively.

We say that the sequence $d[t]$ for $t\geq 0$ is \emph{feasible} if there exists a sequence $u[t]$ for $t\geq 0$ such that, for some constant $C>0$,
  $x_i[t]<C$
for all $t \geq 0$ and for all $i$ ranging over the indices of links in the network. The sequence $d[t]$ is said to be \emph{infeasible} otherwise. In other words, the sequence $d[t]$ is feasible if there exists a control sequence that fully accommodates the exogenous incoming flow. It is straightforward to show that if $x_i[0]\leq \bar{x}$, then $x_i[t]\leq \bar{x}$ for all $t\geq 0$ for any link with an upstream link (\emph{i.e.}, $i\in\{2,3,\ldots,N\}$ for the simple freeway or $i\in\{-M+1,-M+2,\ldots,2N\}$ for the diverging freeway). Thus traffic may only accumulate at onramps or the first link of the network (\emph{i.e.}, link $1$ for the simple freeway and link $-M$ for the diverging freeway), and therefore these are the only links for which it may not be possible to find $C$ satisfying $x_i[t]<C$ for all $t\geq 0$.

 Infeasible exogenous flows are useful for modeling situations in which the freeway network has inadequate capacity, such as during rush hour periods when a large influx of vehicles arrives at the network. Of course, in actual traffic networks, such periods do not last indefinitely, but it may nonetheless be valuable to model such scenarios using an infinite time horizon to provide insights into, \emph{e.g.}, equilibrium conditions that may arise for extended periods of time \cite{Gomes:2008fk}.

For example, \cite{Gomes:2008fk} suggests exogenous flows that are infeasible but on the cusp of feasibility. Specifically, for the simple freeway network, such a choice is
\begin{align}
  \label{eq:2}
d _1[t]&= 40\quad \forall t\\
d_{i}[t]&= 10+\epsilon_i\quad \forall t,\quad i\in\{1',2',\ldots,(N-1)'\}.
\end{align}
When $\epsilon_i=0$ for all $i$, the demand is feasible and the outgoing flow for link $i\in\{1,2,\ldots,N\}$ is 40 vehicles per period at equilibrium. Of these, $30=\beta 40$ advance downstream to link $i+1$, joined by $10$ additional vehicles from link $i'$. However, the flow is infeasible when $\epsilon_i>0$ for some $i$ since the outgoing flow of link $i+1$ cannot exceed $c=40$ vehicles per period.  In the case of infeasible flows, it has been shown that ramp metering can increase throughput of the network \cite{Gomes:2008fk}. Similar choices for exogenous flow can be made for the diverging freeway network. Moreover, it is suggested in \cite{Coogan:2014hl} to model the exogenous flow as belonging to a set, \emph{e.g.}, $d_{i}[t]\in [10-\delta_i,10+\delta_i]$ for some $\delta_i>0$ for all $i\in\{1',2',\ldots,(N-1)'\}$.

\section{Performance Specifications}
\label{sec:perf-spec}
We now suggest several performance metrics that are natural for traffic networks. 

\noindent\textbf{\emph{Total Travel Time}}

The total travel time ($\texttt{TTT}$) of the network up to time $T$ is defined as
\begin{align}
  \texttt{TTT}[T]=\sum_{t=0}^T\sum_{i}x_i[t]
\end{align}
where $i$ is assumed to vary over all state indices of the network. Total travel time is a useful metric when considering finite time horizons for which we seek to minimize the total travel time at the end of the horizon.

\vspace{5pt}
\noindent\textbf{\emph{Throughput}}

Another common performance metric is throughput at time $t$, denoted $W[t]$, and defined as
\begin{align}
W[t]= \sum_{i\in \mathcal{F}} (1-\beta) \min\left\{D(x_i[t]),\frac{\alpha}{\beta} S(x_{i+1}[t])\right\}+\sum_{i\in\mathcal{E}}D(x_i[t])
\end{align}
where

\begin{align}
 \mathcal{F}&=\begin{cases}
        \{1,2,\ldots,N-1\}&\text{ for the simple freeway,}\\
    \{-M,-M+1,\ldots,-1,1,2,N-1,N+1,\ldots,2N-1\}\hspace{-1.7in}\\
&\text{ for the diverging freeway},
  \end{cases}\\
\mathcal{E}& =
  \begin{cases}
    \{N\}&\text{ for the simple freeway,}\\
    \{N,2N\}&\text{ for the diverging freeway.}
  \end{cases}
\end{align}
Throughput is a measure of the number of vehicles that exit the network in a given time step, and total throughput over a horizon $T$ is
\begin{align}
  \label{eq:4}
  J[T]&=\sum_{t=0}^T W[t].
\end{align}
Throughput is a useful metric in cases when the exogenous flow is infeasible or the time horizon is infinite, in which case a discounted or average reward function may be defined as
\begin{align}
  J&=\sum_{t=0}^\infty \gamma^t W[t]\quad \text{or} \quad  J=\limsup_{T\to \infty}\frac{1}{T}\sum_{t=0}^T W[t],
\end{align}
where $\gamma<1$ is a discount factor. Alternatively we may require, for example, that throughput remain above a given threshold for all time.

\vspace{5pt}
\noindent\textbf{\emph{Congestion}}

\newcommand{\xcrit}{x^\text{crit}}
Notice that $D(\xcrit)=S(\xcrit)$ for 
\begin{align}
  \xcrit=\max\left\{\bar{x}-\frac{c}{w},\frac{w\bar{x}}{v+w}\right\}=80
\end{align}
where the second equality is valid for the values reported in Table \ref{tab:params}. Link $i$ is said to be \emph{congested} if $x_i>\xcrit$. %

A third possible performance specification for the system is that $x_i\leq \xcrit$ for all $i$. Motivated by specifications expressible in \emph{temporal logic} \cite{Pnueli:1977ye, Baier:2008vn}, we may require that this conditions holds for all time, or eventually at some time in the future and forever thereafter, or at infinitely many time instants in the future, \emph{etc.} For example, it is possible to characterize a degree of robustness for the system by considering these various possibilities \cite{Tabuada:2015pd}.

\section{Discussion}
\label{sec:discussion}
We now point to some important properties of the traffic model proposed in this note. First, we note that the arguments of the minimization functions that appear in \eqref{eq:dem}, \eqref{eq:m1}--\eqref{eq:m2}, and  \eqref{eq:d1}--\eqref{eq:d3} are all affine so that the model is \emph{piecewise affine} (PWA), that is, there exists a partition $P_1,\ldots,P_M$ of $\mathbb{R}^n$ and collections of matrices $\{A_i\}_{i=1}^M\subset\mathbb{R}^{n\times n}$ and $\{B_i\}_{i=1}^M\subset\mathbb{R}^{n\times m}$ such that
\begin{align}
  x[t+1]=A_ix[t]+B_i u[t]+Ed[t]
\end{align}
whenever $x[t]\in P_i$, that is, the system dynamics are affine in each region of the partition. The matrix $E\in\{0,1\}^{n\times q}$ is a binary matrix with no more than one nonzero entry per row whose role is to ensure that the disturbance vector conforms with the state vector appropriately.

Since the model is piecewise affine, it is, in principle, amenable to tools such as model predictive control \cite{Borrelli:2003fk} and synthesis for linear temporal logic (LTL) objectives \cite{Yordanov:2012fk} which have been specialized to PWA systems. The latter approach is considered in \cite{Coogan:2014pi} to meet LTL specifications.

Next, it has been shown in \cite{Gomes:2008fk, coogan2015compartmental, Lovisari:2014qv} that traffic networks with only merging junctions are \emph{monotone} dynamical systems for which trajectories maintain a partial order on states \cite{Hirsch:2005ek, Angeli:2003fv}. Moreover, general traffic networks with diverging junctions possess dynamics that are mixed monotone \cite{Coogan:2016rp}, a generalization of monotone systems \cite{Smith:2008rr}. For such systems, reachable sets are over-approximated by evaluating a certain decomposition function, derived from the dynamics, at only two extremal points, regardless of the state-space dimension. Thus, such systems are amenable to efficient finite abstraction \cite{Coogan:2014ty}, a typical requisite for applying formal methods techniques to control systems \cite{tabuada2009verification}. This efficiency is exploited in \cite{Coogan:2014hl} to synthesize control strategies for traffic networks. Additionally, it is shown in \cite{Kim:2016et} that monotone systems are especially amenable to classes of temporal logic specifications that encourage lower occupancies.

Moreover, the sparse interconnection structure of many traffic networks suggest compositional techniques for further aiding scalability \cite{Kim:2015ff}. Finally, some works have suggested that scalability can be improved by avoiding discretization of the state space; in particular, \cite{Le-Corronc:2013fk} proposes computing abstraction based on sequences of applied control inputs, and this approach is applied to a traffic flow model similar to that presented here. Nonetheless, these formal approaches are only applicable to relatively small networks (up to approximately 10 state dimensions).

\section{Conclusions}
\label{sec:conclusions}
We have presented a general, hybrid model of traffic flow in vehicular transportation networks and suggested two simple classes of networks that encompass realistic scenarios. Both classes are easily scaled to allow networks of arbitrary size.  We have further characterized several performance metrics that are practically motivated and fit well into optimization or formal methods frameworks. Traffic flow models have already proven to be effective case studies for new tools in hybrid systems, and it is our hope that the transportation network model detailed in this note may offer a practical and extensible benchmark problem for applying new tools and techniques for hybrid systems.

\section{Acknowledgments}
Research funded in part by the National Science Foundation under grant 1446145.

\bibliographystyle{ieeetr}
\bibliography{$HOME/Documents/Books/books}

\end{document}